\documentclass{aa}
\usepackage{graphicx}
\usepackage{txfonts}
\usepackage{rotating}
\usepackage{color}
\usepackage{url}
\usepackage{natbib}
\usepackage{float}
\usepackage{textcomp,gensymb}

\usepackage[utf8]{inputenc}

\usepackage{amsmath} 

\makeatother

\begin{document} 

 \title{A progenitor candidate for the type II-P supernova SN~2018aoq in NGC~4151}
   \author{D. O'Neill\inst{1}
   \and R. Kotak\inst{2}
   \and M. Fraser\inst{3}
   \and S.~A. Sim\inst{1}
   \and S. Benetti\inst{4}
   \and S.~J. Smartt\inst{1}
   \and S. Mattila\inst{2}
   \and C. Ashall\inst{5}
   \and E. Callis\inst{3}
   \and N. Elias-Rosa\inst{6,7}
   \and M. Gromadzki\inst{8}
   \and S.~J. Prentice\inst{1}
}  
        \institute{Astrophysics Research Centre, School of Mathematics and Physics, Queen's University Belfast, Belfast, BT7 1NN, UK. \\ (\email{doneill955@qub.ac.uk} \label{inst1})
    \and Department of Physics and Astronomy, Vesilinnantie 5, University of Turku, 
    FI-20014, Turku, Finland. \label{inst2}
    \and School of Physics, O'Brien Centre for Science North, University College Dublin, Belfield, Dublin 4, Ireland. \label{inst3}
    \and INAF - Osservatorio Astronomico di Padova, Vicolo dell’Osservatorio 5, 35122 Padova, Italy.\label{inst4}
   \and Department of Physics, Florida State University, Tallahassee, FL 32306, USA.\label{inst5}
   \and Institute of Space Sciences (ICE, CSIC), Campus UAB, Carrer de Can Magrans s/n, 08193 Barcelona, Spain.\label{inst6}
   \and Institut d’Estudis Espacials de Catalunya (IEEC), c/Gran Capit\'a 2-4, Edif. Nexus 201, 08034 Barcelona, Spain.\label{inst7}
   \and Warsaw University Astronomical Observatory, Al. Ujazdowskie 4, 00-478 Warszawa, Poland.\label{inst8}
}

    \abstract{
We present our findings based on pre- and post-explosion data of the type II-Plateau SN~2018aoq that exploded in NGC~4151. As distance estimates to NGC~4151 vary by an order of magnitude, we utilised the well-known correlation between ejecta velocity and plateau brightness, i.e. the standard candle method, to obtain a distance of 18.2$\pm$1.2\,Mpc, which is in very good agreement with measurements based on geometric methods. The above distance implies a mid-plateau absolute magnitude of $M_{V}^{50}=-15.76\pm$0.14 suggesting that it is of intermediate brightness when compared to IIP SNe such as SN~2005cs at the faint end, and more typical events such as SN~1999em. This is further supported by relatively low  expansion velocities (\ion{Fe}{ii} $\lambda$5169 $\sim$3000\,km\,s$^{-1}$ at +42~d). Using archival {\it HST}/WFC3 imaging data, we find a point source coincident with the supernova position in the F350LP, F555W, F814W, and F160W filters. This source shows no significant variability over the $\sim$2\,month time span of the data. From fits to the spectral energy distribution of the candidate progenitor, we find 
$\log\left(L/L_\odot\right)\sim 4.7$ and $T_{\mathrm{eff}}\sim 3.5$\,kK, implying an M-type red supergiant progenitor. From comparisons to single and binary star models, we find that both favour the explosion of a star with a zero-age main sequence mass of $\sim$$10\,M_\odot$. 

}
     

\maketitle

\section{Introduction}

\noindent

Although type II-plateau\,(P) supernovae (SNe) are the most commonly occurring subtype of explosions resulting from the core-collapse (CC) of massive evolved stars \citep{LOSSrate}, a number of issues remain
to be explored relating to their evolution prior to collapse and the origin of diversity in their observed properties. Direct detections of sources at the SN site in pre-explosion imaging have led to a growing body of evidence in favour of red supergiants (RSG) being the progenitors of type IIP SNe \citep{03gdVD,03gdpro}. 
Based on the consideration of a sample of 20 IIP SNe within 25 Mpc and a combination of detections and limits to the presence of a point source at the SN location, the progenitors have inferred masses in the range $8.5^{+1.0}_{-1.5} \lesssim M_{\odot} \lesssim 16.5^{+1.5}_{-1.5}$  \citep{RSGP}. 

There is considerable diversity in the observed photometric and spectroscopic behaviour of type IIP SNe. At the faint end of the IIP brightness distribution ($-13\lesssim M_V\lesssim -15$),
it is arguably natural to assume that they might simply originate from stars that are only just massive enough to undergo Fe CC. Indeed, there is some evidence to support this view from progenitor detections, for example SNe 2005cs and 2008bk \citep{05cspro,08bkMass,2008bkVLT,08bkpro}. However, it is plausible that faint SNe arising from CC due to electron captures on $^{24}$Mg and $^{20}$Ne in the cores of super asymptotic giant branch stars \citep{ECSNeNomoto} also play a role. Uncertainties in modelling processes such as mass loss, mixing, and convection together with the influence of metallicity have resulted in a wide mass range ($7-11\,M_\odot$) over which electron-capture SNe can occur. 
Although the picture is likely to be far more complicated than described above, gathering progenitor masses estimated from multi-band pre-explosion imaging, provides valuable information against which models can be anchored.\\
SN~2018aoq (Kait-18P) was discovered on 2018 April 01.436 by the Lick Observatory Supernova Search (LOSS) at +15.3 mag unfiltered magnitude \citep{DiscAtel} on the edge of the intermediate spiral Seyfert galaxy NGC~4151 (Fig. \ref{image}). It was spectroscopically classified as a type II SN on 2018 April 02 \citep{18aoqClass}.
Based on early indications that it might be fainter than the typical type IIP population, and motivated by the abundance of pre-explosion imaging of NGC~4151, we initiated a follow-up campaign as described below.

\section{Observations and data reduction}

Optical imaging data were collected primarily using a combination of ATLAS
\citep[Asteroid Terrestrial-impact Last Alert System;][]{ATLASton} and the 2.0 m Liverpool Telescope (LT). Spectroscopic follow-up was carried out primarily at the  2.5m Nordic Optical Telescope (NOT) as part of the NOT Unbiased Transient Survey (NUTS\footnote{\url{http://csp2.lco.cl/not/}}). A complete record of photometric and spectroscopic data can be found in tables \ref{phot}, \ref{ATLAStable}, and \ref{spectable}. 
Standard procedures of 
bias subtraction, trimming the overscan regions, and flat-field correction were applied to the raw imaging and spectroscopic data.
Instrumental magnitudes for the SN photometry were found by fitting a point spread function (PSF) with the SNOoPy \textsc{IRAF} package\footnote{\url{http://sngroup.oapd.inaf.it/snoopy.html} } to the reference stars listed in table \ref{Cal} and subtracting this PSF from the SN. The photometric zero point corrections for both the LT and NOT images  were calculated using the instrumental magnitudes of the stars in the field (table \ref{Cal}) and comparing these to the magnitude values given by SSDS DR12 \citep{SDSS12} in each of the filters. In the case of the $B$ and $V$ bands, the magnitudes in the Sloan bands were transformed using the transformations listed in \cite{SDSSJester}. 
For the spectroscopic data, the wavelength calibration was carried out using He-Ne arcs, and relative flux calibration was performed using spectrophotometric standards taken on the same night, and with the same instrumental configuration as for SN~2018aoq. Absolute flux calibration of the spectra was achieved using the broadband imaging points interpolated to the epochs of spectroscopy. From the  ATLAS non-detection on 2018 March 28 and the discovery on 2018 April 01 (figure \ref{lc2}, Table \ref{ATLAStable}), we adopt 2018 March 30 (MJD\,=\,58208) as the explosion date.

\section{Reddening and distance}
\label{sec:rd}

\begin{figure}[!t]
  \centering
   \includegraphics[width=\hsize]{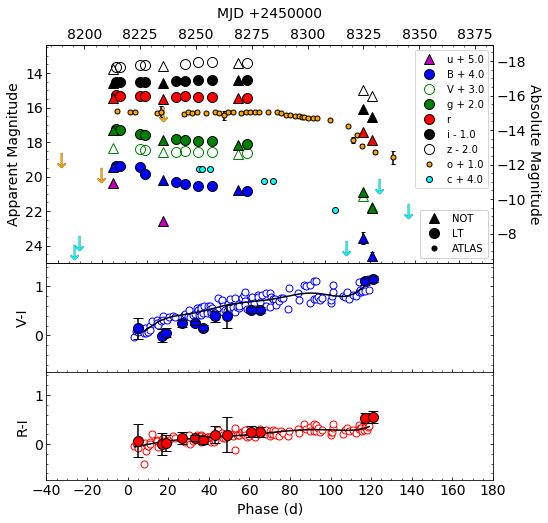}
      \caption{\textbf{Top panel:} Light curves of SN~2018aoq in various filters. Upper limits are indicated by downward pointing arrows. The letters `c' and `o' indicate the `cyan'  ($\lambda_{\mathrm{eff}}\sim$5330\AA) and `orange' ($\lambda_{\mathrm{eff}}\sim$6790\AA) ATLAS filters, respectively. The non-detection ($>$18.85\,mag) on 2018 March 28 (MJD\,=\,58206) allows us to constrain the rise time to $\lesssim$5\,d. 
      \textbf{Bottom panel:} Extinction-corrected $V-I$ and $R-I$ colour evolution for SN~2018aoq. The errors are the result of compounding errors in transforming from the $V$, $r$, and $i$ filters to Johnson $R$ and $I$. Open circles are extinction-corrected values taken from 8 well-sampled type IIP SNe (table \ref{ExSample}). The black line indicates a fit to these data from which we infer $E(B-V)=0.04$  for SN~2018aoq. }
         \label{lc2}
\end{figure}

The Galactic reddening in the line of sight to SN~2018aoq is found to be $E(B-V)=0.02$ \citep{SF}. We attempt to estimate the host galaxy reddening by comparing the colour evolution and spectral shape to other type IIP SNe.
Figure \ref{lc2} shows the $V-I$ and $R-I$ colour evolution of SN~2018aoq compared to a sample of well-observed IIP SNe (Table \ref{ExSample}). 
A fairly uniform colour evolution is expected over the first $\sim$100\,d as hydrogen recombination dominates \citep{Eastman96}. 
A least squares fit using low-order polynomials was carried out on the 
extinction-corrected colour curves of the sample SNe. The $V-I$ and $R-I$ points of SN~2018aoq were then reddened by iteratively varying the $E(B-V)$ value until the best fit was obtained.
The colour evolution of SN~2018aoq tracks the general IIP evolution; a best fit is obtained for the total host and Galactic extinction $E(B-V)_{\mathrm{total}}\sim0.04$. The $V-I$ colour appears to be systematically bluer than the sample, but still within the range exhibited by other IIP SNe.

Spectra of SN~2018aoq were also compared to the extinction-corrected spectra of SNe~1999em and 2005cs \citep[with E(B-V) values of 0.1 and 0.05 mag, respectively;][]{99em,Pa05cs} at similar epochs. We reddened the SN~2018aoq spectra incrementally until a reasonable match was obtained; we found this value to be $E(B-V)\sim0.03$. Also, we do not clearly detect any feature attributable to the Na I D lines in our earliest spectrum.  

We further note that the location of the SN in a relatively isolated region at 
$\sim$73\arcsec\ from the nucleus of the host galaxy is in line with the low inferred extinction. Based on the findings above, we conclude that the SN suffers from low extinction and adopt of a value of $E(B-V)_{\mathrm{tot}}=0.04$ mag in what follows.

There is a great variation in the estimated distances to NGC~4151, ranging from 3.8--29.2\,Mpc as listed on NED\footnote{\url{https://ned.ipac.caltech.edu/}}.
NGC~4151 has received considerable attention over the years as it is one of the very few local AGN for which dynamical mass estimates of the central black hole can be compared to those obtained from reverberation mapping and other methods, thereby providing a calibration of the black hole mass scale. However, anchoring the calibration to NGC~4151 requires a reliable distance to it. Using the heliocentric radial velocity of the galaxy is 998\,km\,s$^{-1}$ \citep{pedlar92} and yields a distance of 14.1\,Mpc with $H_0=71$\,km\,s$^{-1}$\,Mpc$^{-1}$. However, strong peculiar motions render this value unreliable.  Distance estimates using the Tully-Fisher method do not fare much better as the orientation of NGC~4151 is almost perpendicular to our line of sight, making measurements of the rotational velocity challenging. Based on a variant of the quasar parallax method proposed by \citet{QuasarPrlx} to obtain direct distances to quasars, \cite{4151dist} reported a value of 19$^{+2.4}_{-2.6}$\,Mpc. This value is reliable as it depends on purely geometrical arguments rather than requiring an underlying source of known luminosity.

In order to obtain an independent estimate for the distance, we applied the standard candle method (SCM) as described in \cite{SCM}.
Essentially, the SCM is the manifestation of a positive correlation between the luminosity and expansion velocity during the plateau phase; a convenient reference point is taken to be at mid-plateau ($\sim$50\,d). 
Fortuitously, we have $V$ and $i$-band measurements taken on May 18, 49\,d post explosion (Table \ref{phot}); to convert to the $I$-band, we used the transformations in \cite{SDSSJester} to find a value of $I = 15.25\pm0.29$ mag. The \ion{Fe}{II} $\lambda$5169 velocity at 50\,d was found by linear interpolation between the velocity values found at +42\,d and +61\,d, which yielded a value of 2699$\pm$120 km\,s$^{-1}$. We used the sample provided by \citet{Joe} consisting of type IIP SNe that occurred in galaxies with Cepheid distances for calibration. 

We find $D_I$ = 18.9$\pm$1.5\,Mpc and $D_{V}$ = 17.3$\pm$1.8\,Mpc with a weighted average of 18.2$\pm$1.2\,Mpc, which is the value we adopt in this work. We note that this value is in excellent agreement with the value based on direct methods \citep{4151dist}.

\section{Supernova photometry and spectroscopy}

Using the values for the distance and reddening from \S\,\ref{sec:rd}, we find $M_{V}=-15.76\pm$0.14 at an epoch of 49\,d, i.e. approximately mid-plateau. 
Thus, SN~2018aoq appears to lie at a brightness intermediate between the normal and subluminous IIP populations, although the demarcations in brightness between these groups are not strictly defined. 
We note in passing that the rise time to the plateau was relatively rapid ($\lesssim$5\,d; figure \ref{lc2}), in keeping with previous studies, for example \citet{10id} and \citet{EEErise}, who reported a possible correlation between rise time and absolute peak brightness.

From a blackbody fit to the photometric data at +49\,d, we estimate a quasi-bolometric luminosity of $L_{\mathrm{Bol}}=$10$^{41.8}$\,erg\,s$^{-1}$. Not surprisingly, this is fainter than that of SN~1999em, a normal II-P and marginally brighter than SN~2005cs, $L_{\mathrm{Bol}}=$10$^{41.4-41.7}$\,erg\,s$^{-1}$ \citep{IIPbolo,Pa05cs}. 
The photometric data (Tables  \ref{phot}, \ref{ATLAStable}) allow us to constrain the plateau duration to $\sim$80\,d following which the light curve drops at a rate of 0.02~mag\,day$^{-1}$ ($o$-band) for $\sim$20\,d, when it abruptly drops onto the nebular tail phase. 
This drop is $\sim$2\,mag compared to the mid-plateau brightness and is shallower than that of SN~2005cs in a comparable filter \citep[m$_R$=2.7~mag ][]{Pa05cs}.

The 4--111\,d spectral evolution of SN~2018aoq is shown in figure \ref{spectra}.
The spectra are typical of type IIP SNe; the early spectra are blue, which supports our finding (cf. \S \ref{sec:rd}) of minimal extinction. The Balmer lines and \ion{He}{i} $\lambda$5875 are clearly discernible in the spectra. 
By 42\,d, the metal lines for instance \ion{Fe}{ii} $\lambda$5169, \ion{Ba}{ii} $\lambda$6142, and \ion{Sc}{ii} $\lambda$6246 have all emerged, as has the \ion{Ca}{II} triplet. By 61\,d, the forbidden [\ion{Ca}{II}] $\lambda$7291 emerges -- all of these features are expected to be present in type IIP SN spectra at these epochs. From the absorption minimum in the 4\,d spectrum, we measure an H{$\alpha$} velocity of $\sim$9700\,km\,s$^{-1}$, which is $\sim$2600\,km\,s$^{-1}$ slower than the corresponding velocity measured for the earliest spectra available for SN~1999em \citep[+5\,d,][]{99emEPM}.
Likewise, the velocities measured from the metal lines in the 42\,d spectrum are $\sim$800\,km\,s$^{-1}$ lower than those of typical IIP SNe such as SN~1999em \citep{99em} at similar epochs, but $\sim$800\,km\,s$^{-1}$ higher than those from the low-luminosity group such as SN2005cs \citep{Pa05cs}: 2960$\pm$60\,km\,s$^{-1}$ and 2689$\pm$60\,km\,s$^{-1}$, measured for
\ion{Fe}{ii} $\lambda$5196 and \ion{Sc}{ii} $\lambda$6246, respectively. A spectral comparison with SNe~1999em and 2005cs that illustrates the above is shown in figure \ref{specComp}.

\section{Archival data and progenitor identification}
\label{sec:archival}

\begin{figure}[t]
 \hspace*{0.1cm}
   \includegraphics[width=\hsize]{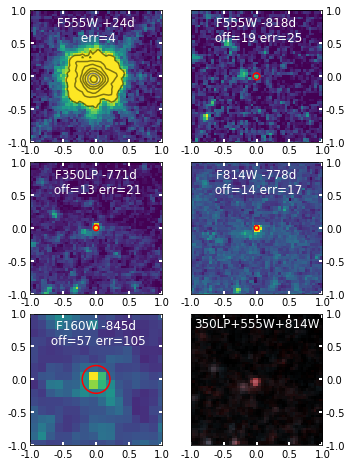}
      \caption{Top left panel: "$2''\times2''$ {\it HST\/}/WFC3 image stamps. The top left panel shows SN~2018aoq  at an epoch of 24\,d after explosion while all others show pre-explosion images (Table \ref{Archive}). The filter and epoch are labelled. The offset between the star and SN position, and the associated positional error (in mas) is also indicated. The centroid of the SN position, as measured from the image in the top left panel, is shown with a red circle, the radius of which is twice the uncertainty in the position for clarity. 
      The limiting and absolute magnitudes for the progenitor in each filter are m$_{F555W}$>28.0 ($-$3.3),  m$_{F350LP}$>26.8 ($-$4.5), m$_{F814W}$>26.0 ($-$5.3), and m$_{F160W}$>24.1 ($-$7.2).
      }

         \label{Detections}
   \end{figure}
A wealth of pre-explosion archival data are available for NGC~4151. We focus primarily on pre-explosion WFC3 imaging data from the {\it Hubble Space Telescope (HST)} taken approximately 2\,yr prior to the occurrence of SN~2018aoq (Table \ref{Archive}).  
Luckily, a post-explosion image (+24\,d) in the F555W filter also taken with the WFC3 instrument was publicly available,\footnote{HST Proposal ID:15151} allowing us to accurately map the position of the SN to pre-explosion images. 
The position of the SN in the pre-explosion drizzled images was found using a linear transformation computed within the geomap and geoxytran tasks in IRAF using five to six field stars in common. Although the field of view of the WFC3 instrument is $162''\times162''$, the difference in exposure times of the pre- and post-explosion F555W images meant that only a few stars were available in the two F555W images to calculate the coordinate transformation. However, as previously noted, SN~2018aoq does not lie in a crowded field. 

Fig. \ref{Detections} shows the result of the above procedure in various filters. The 3$\sigma$ detection limit in each filter was found by adding artificial stars created using TinyTim\footnote{\url{http://www.stsci.edu/hst/observatory/focus/TinyTim}}. The artificial stars were then scaled such that the peak count of the PSF matched the background noise + 3$\sigma$. Photometry on these artificial stars was then carried out as previously described. The limits are given in Fig. \ref{Detections}.
A bright point source was identified in the images very close to the SN position. For the F555W pre- and post-explosion images, the offset between the SN centroid and the point source was found to be 19~mas.

In order to robustly estimate the errors, we ran Monte Carlo simulations on synthetic data.
The position of the peak was allowed to vary from $-$100 mas to +100 mas in both directions in steps of 0.4\,mas. The process was repeated for 1000 iterations. The discrepancy between this position and that found previously is taken to be representative of the error.
The total error in position is the combination, in quadrature, of the positional errors in the SN and pre-explosion source, as well as the error in the transformation. For the F555W filter (available for both pre- and post-explosion epochs), we find the error to be 25\,mas. For all filters, we find the offsets to be within the total error estimate (Fig. \ref{Detections}), and therefore conclude that the source is coincident with SN~2018aoq. 

Photometry on the star was carried out using DOLPHOT\footnote{\url{http://americano.dolphinsim.com/dolphot/}}. The measurements for the star in all available filters are listed in Table \ref{Archive} and shown in figure \ref{PrgLc};  The progenitor star had a mean absolute magnitude in the optical bands of $M_{F350LP}=-5.69$, $M_{F555W}=-4.74$, $M_{F814W}=-7.39$ and in the near-infrared: $M_{F160W}=-9.48$. The pre-explosion photometry covers a period of 73\,d over the course of which  there appears to be little or no variation within a few tenths of a magnitude in any of the filters. 

\begin{figure}
   \centering
   \includegraphics[width=\hsize]{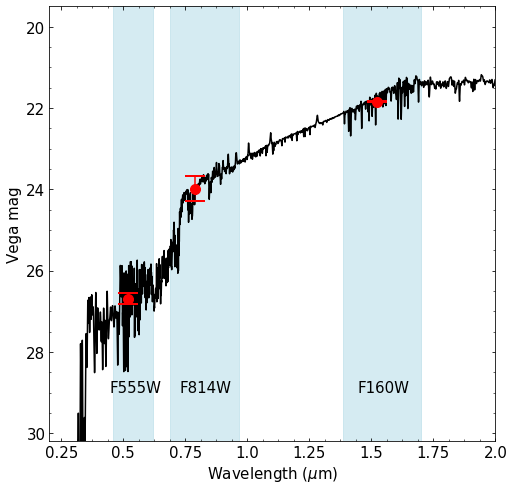}
      \caption{Best-fitting synthetic SED from the PHOENIX stellar atmosphere models corresponding to  spectral-type M with T=$3500$\,K, log~g=0.0 and $E(B-V)=0.03$, in units of Vega mag. The average scatter between the SED and observed {\it HST} photometry $\sigma$=0.05\,mag. The shaded regions indicate the wavelength range of each of the filters. The red points denote the average measured magnitude in each of the {\it HST} filters. 
      }
         \label{SEDmag}
   \end{figure}

 \begin{figure*}[!t]
   \centering
    \includegraphics[height=0.5\textwidth]{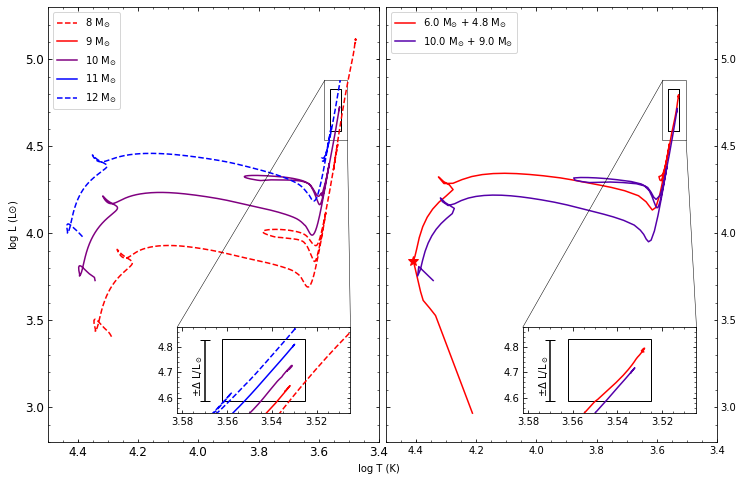}
      \caption{ Evolutionary tracks computed at solar metallicity ($Z=0.02$) and selected to satisfy the inferred temperature and luminosity of progenitor as described in \S\,\ref{sec:archival} and delineated by the box.
 \textbf{Left:} Single star models from the STARS code. 9-11\,$M_\odot$ models matched the progenitor properties at its endpoint. The 8\,$M_\odot$ and 12\,$M_\odot$ tracks are shown for comparison. The inset shows a zoom-in of a finer grid of models with masses as indicated in the legend.
 \textbf{Right:} Evolutionary tracks resulting from binary evolution computed using the BPASS code. In total, 253 tracks
 matched the progenitor properties. These had primary stars in the mass range $6-10$\,$M_\odot$, and 50\% had a 10\,$M_\odot$ primary. Shown here is a 10\,$M_\odot$ primary with a 9\,$M_\odot$ secondary on a long period orbit ($\sim$4000~d) with little interaction (blue) and an early merger between stars in a small orbit($\sim$10~d), which initially consisted of a 6\,$M_\odot$ primary and 4.8\,$M_\odot$ secondary resulting in a star with a mass between 10\,$M_\odot$ and 11\,$M_\odot$ (red). The red star indicates the position on the isochrone at which the merger was completed.}
  \label{iso}
\end{figure*}

In order to characterise its properties, we used the HST synthetic photometry package stsynphot.\footnote{\url{https://stsynphot.readthedocs.io/en/latest/}}
From the measured average apparent magnitude in each filter, we generated a spectral energy distribution (SED) using models taken from the PHOENIX stellar atlas \citep{phoenix}. Using the prescriptions provided by \cite{12awdust} for circumstellar dust, we simultaneously varied the temperature, extinction, log surface gravity (g), and optical depth ($\tau_V$) of the circumstellar dust to generate synthetic magnitudes in the {\it HST} filters and compared these with our measured values. The best-fitting parameters, i.e. those that returned the smallest mean scatter ($\sigma$=0.056 mag), were found to be a temperature of $T=3500\pm$150\,K, $\log(g)=0.0$ and minimal circumstellar extinction $\tau_V=0.0$. The resulting SED is shown in Fig. \ref{SEDmag}.

For completeness, we note that no source is visible in pre-explosion imaging at 3.6 and 4.8 $\mu$m in Spitzer/IRAC imaging down to limiting Vega magnitudes of >20.7 and >20.1 (from artificial star tests), respectively; these are consistent with the SED.

Integrating over the SED allows us to obtain the total flux from the progenitor, which can then be used to constrain the luminosity and radius. 
In order to explore a conservative parameter range, we folded into the luminosity bounds the entire range of SCM distances implied by the uncertainties (i.e. 15.5 - 20.4\,Mpc) in \S\,\ref{sec:rd}.
We found a luminosity range of $\log\left(L/L_\odot\right)= 4.59-4.83$ (4.72$\pm$0.12) for the progenitor candidate, making it a likely red supergiant of spectral type M.
To estimate its initial mass, we considered evolutionary tracks computed assuming solar metallicity for both single- and binary-star models that contain similar treatments of the dominant physical processes \citep[STARS, BPASSv2.1,][ respectively]{STARS2,BPASS2-1}. Models were selected by comparing the inferred luminosity and temperature to the calculated values. 

For the single star case, three models matched the progenitor properties with masses in the 9$-$11\,$M_\odot$ range (figure \ref{iso} left panel).
A comparison with the non-rotating Geneva models \citep{Meynet}, based on a slightly coarser mass grid, is consistent with the above, yielding a likely progenitor in the range  $8\leq M_{\odot}\leq12$. 
We found 253 binary models that matched the progenitor properties out of $\approx$12600 BPASS models considered. 
The vast majority ($\sim$80\%) of these consisted of a 9$-$11\,$M_\odot$ star in a long period orbit ($P\gtrsim1000$~days) with a range of companion masses (1\,$M_\odot$ -- 10\,$M_\odot$). The wide orbits mean minimal mass transfer between the primary and secondary stars during the course of their respective evolutionary phases. Thus, the mass of the companion is of little consequence. The remaining models result in mergers between two lower mass stars in tight orbits with $P\lesssim$10~days, resulting in a star with a mass close to 10-11\,$M_\odot$. Thus, both the single- and binary star models favour a star in the range 10$\pm$2$\,M_{\odot}$. There is still some uncertainty regarding the temperature range of RSG stars and the methods used to derive them \citep{RSGDav,RSGLev}. In order to ensure that our results are not highly dependent on the estimated temperature, we increased the temperature range considered to $2500-4500$\,K and repeated the process.
We found the results to be entirely consistent with the above.

From the {\it HST} photometry and SED (figure \ref{SEDmag}), we can rule out a bright companion star. A single SED matches the HST photometry very well, so a companion would have to have a SED that either peaks sharply in the ultraviolet region or in the mid-infrared. The latter is unlikely, given the non-detections in the {\it Spitzer}/IRAC bands. 
In order to determine the robustness of the above statements, we combined the progenitor SED with that obtained from synthetic spectra with $2000 \lesssim T_{\mathrm{eff}} \lesssim 30000$ and $10 \lesssim $ $m_{F814W}$ $ \lesssim 30$ to examine whether it might possible for a companion to exist alongside the progenitor, but without affecting the SED (Fig. \ref{SEDmag}). This exercise also allows us to assess the feasibility of detecting a companion once the SN has faded away.
We were unable to find such a case within the above range of parameters as any companion star would be fainter than $M_{F814W}$=$-1.3$).

Therefore, we conclude that the progenitor was a single star with $M_{ZAMS}$=10$\pm$2\,$M_{\odot}$, in a wide binary with a low mass companion, or that it resulted from the merger of two low mass stars with a combined mass close to $M_{ZAMS}$=10$\pm$2\,$M_{\odot}$.  
In either of the above cases, a clear prediction would be that very late time imaging observations should show that the source we identified as the progenitor has disappeared because a putative companion would be too faint to detect. Similar results based on an in-depth study of model dependencies related to type II SNe has been found by \cite{Zap}.

\section{Summary}
We have presented the case for the detection of the progenitor to the II-P SN~2018aoq. Using pre-explosion data from {\it HST}, we find that the SED best matches a RSG star that has a temperature of $\sim 3500\pm$150\,K and luminosity of $\log\left(L/L_\odot\right)\sim 4.7$, has minimal Galactic and circumstellar dust extinction, and displays minimal variability in the lead up to CC. Both single and binary star models favour a progenitor with a ZAMS mass of $10\pm2\,M_{\odot}$. The minimal extinction and isolated location of SN~2018aoq lends confidence to the progenitor detection, making it one of only a few for which multi-filter, multi-epoch observations were possible. 
Although further observations will be required to constrain the $^{56}$Ni mass and to confirm the disappearance of the progenitor, our result adds to the small but growing body of evidence in favour of the explosion of low mass RSGs leading to faint type IIP SNe. Although previous studies have explored the possibility of a link between progenitor mass and explosion energy \citep[e.g.][]{Poznanski13}, from an observational standpoint, the existence (or otherwise) of such a correlation will have to await a larger sample of robust progenitor detections.

\begin{acknowledgements}
We acknowledge comments from an anonymous referee.
We thank JJ Eldridge for patiently answering questions related to BPASS.
D. O'Neill acknowledges a DEL studentship award.
MF acknowledges the support of a Royal Society - Science Foundation Ireland University Research Fellowship.
This work is based on observations made with the Nordic Optical Telescope, operated by the Nordic Optical Telescope Scientific Association at the Observatorio del Roque de los Muchachos, La Palma, Spain, of the Instituto de Astrofisica de Canarias.
The Liverpool Telescope is operated on the island of La Palma by Liverpool John Moores University in the Spanish Observatorio del Roque de los Muchachos of the Instituto de Astrofisica de Canarias with financial support from the UK Science and Technology Facilities Council. The observations were made as part of the JL18A10b programme.
Some of the data presented in this paper were obtained from the Mikulski Archive for Space Telescopes (MAST). STScI is operated by the Association of Universities for Research in Astronomy, Inc., under NASA contract NAS5-26555.
This work is also based on data from HST proposal: 15151 (S.~V. Dyk) retrieved from the MAST archive.
This work has made use of data from the Asteroid Terrestrial-impact Last Alert System (ATLAS) project. ATLAS is primarily funded to search for near-Earth asteroids through NASA grants NN12AR55G, 80NSSC18K0284, and 80NSSC18K1575; byproducts of the NEO search include images and catalogues from the survey area.  The ATLAS science products have been made possible through the contributions of the University of Hawaii Institute for Astronomy, the Queen's University Belfast, and the Space Telescope Science Institute.
SJS acknowledges funding from STFC Grants  ST/P000312/1 and ST/N002520/1.
N.E-R. acknowledges support from the Spanish MICINN grant ESP2017-82674-R and FEDER funds.
MG is supported by the Polish National Science Centre grant OPUS 2015/17/B/ST9/03167.

\end{acknowledgements}

\bibliographystyle{aa}
\bibliography{2018aoq}

\begin{appendix}
\section{Additional data and figures}

\begin{figure}[!ht]
   \centering
   \includegraphics[width=\hsize]{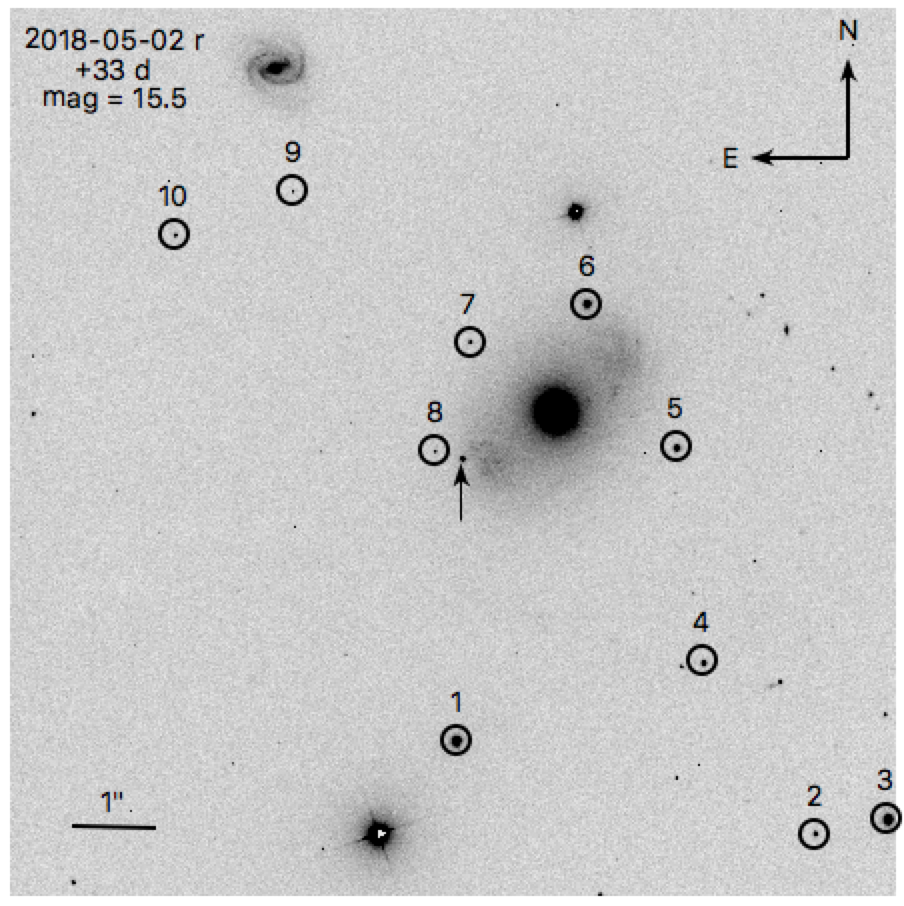}
      \caption{LT IO:O $r$-band image showing the location of SN~2018aoq indicated by the arrow. It is located at $\alpha_{J2000}=12^{h}10^{m}38^{s}.190$, $\delta_{J2000} = +39^{\circ}23'47\farcs00$, or 65\farcs0\,E and 33\farcs6\,S of the nucleus of NGC~4151. The reference stars used for calibration are encircled and listed in Table \ref{Cal}}.
         \label{image}
   \end{figure}
   
        \begin{figure}[!t]
   \centering
   \includegraphics[width=\hsize]{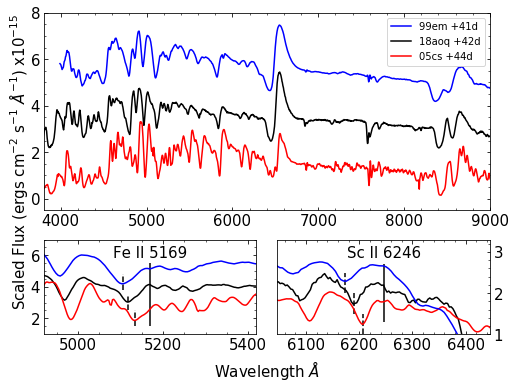}
      \caption{Spectral comparison of SN~2018aoq alongside a normal IIP (SN 1999em) and a low luminosity IIP (SN 2005cs). The bottom panels show the \ion{Fe}{II} and \ion{Sc}{II} absorption in more detail. Dashed lines indicate the minimum of the absorption feature. The solid line indicates the rest wavelength of the feature.}
         \label{specComp}
   \end{figure}
   
   \begin{figure}[!t]
   \centering
   \includegraphics[width=\hsize]{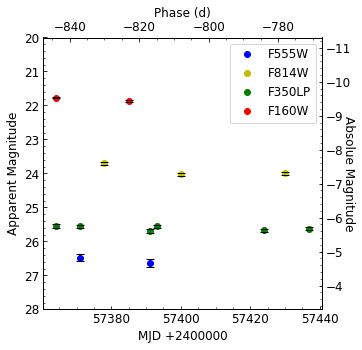}
      \caption{Extinction corrected light curves of the progenitor candidate.}
         \label{PrgLc}
   \end{figure}

\begin{sidewaystable}
\begin{tabular}{c c c c c c c c c c c}        
\hline\hline                 
\tabularnewline[-0.25cm]
Date & MJD & Epoch & u & B & V & g & r &  i & z &Telescope\\
& (+2400000) & (d) &&&&&&\tabularnewline
\hline                        
\tabularnewline[-0.25cm]

2018 Apr 05 & 58213.10 & 5 & 15.57 (0.05) & 15.60  (0.03) & 15.48  (0.03) & 15.44  (0.03) & 15.53  (0.16) & 15.64  (0.04) & 15.82  (0.06) & NOT \\
2018 Apr 06 & 58214.00 & 6 & - & 15.58  (0.03) & - & 15.42  (0.05) & 15.40  (0.10) & 15.58  (0.03) & 15.70  (0.07) & LT \\
2018 Apr 08 & 58216.03 & 8 & - & 15.58  (0.03) & - & 15.48  (0.07) & 15.41  (0.06) & 15.59  (0.06) & 15.68  (0.06) & LT \\
2018 Apr 16 & 58224.97 & 17 & - & 15.60  (0.02) & 15.52  (0.04) & 15.69  (0.07) & 15.42  (0.06) & 15.60  (0.06) & 15.60  (0.01) & LT \\
2018 Apr 18 & 58226.98 & 19 & - & 16.01  (0.03) & 15.60  (0.04) & 15.75  (0.04) & 15.43  (0.03) & 15.58  (0.05) & 15.61  (0.03) & LT \\
2018 Apr 27 & 58235.05 & 27 & 17.75  (0.04) & 16.38  (0.02) & 15.71  (0.03) & 16.00  (0.03) & 15.58  (0.02) & 15.64  (0.04) & 15.64  (0.04) & NOT \\
2018 May 02 & 58240.96 & 33 & - & 16.48  (0.03) & 15.68  (0.03) & 15.98  (0.06) & 15.50  (0.00) & 15.56  (0.01) & - & LT \\
2018 May 06 & 58244.95 & 37 & - & 16.59  (0.03) & 15.67  (0.02) & 16.04  (0.06) & 15.41  (0.03) & 15.51  (0.00) & 15.53  (0.05) & LT \\
2018 May 12 & 58250.97 & 43 & - & 16.69  (0.02) & 15.69  (0.02) & 16.09  (0.03) & 15.50  (0.07) & 15.49  (0.03) & 15.42  (0.04) & LT \\
2018 May 18 & 58256.93 & 49 & - & 16.73  (0.03) & 15.70  (0.02) & 16.11  (0.06) & 15.48  (0.20) & 15.47  (0.03) & 15.42  (0.04) & LT \\
2018 May 30 & 58268.92 & 61 & - & 16.93  (0.02) & 15.80  (0.02) & 16.30  (0.02) & 15.57  (0.02) & 15.50  (0.03) & 15.46  (0.03) & NOT \\
2018 Jun 03 & 58272.96 & 65 & - & 16.98  (0.03) & 15.77  (0.03) & 16.28  (0.05) & 15.53  (0.02) & 15.46  (0.02) & 15.45  (0.06) & LT \\
2018 Jul 25 & 58324.90 & 117 & - & 19.73  (0.34) & 18.23  (0.03) & 19.03  (0.03) & 17.51  (0.02) & 17.16  (0.02) & 17.02  (0.03) & NOT \\
2018 Jul 29 & 58328.88 & 121 & - & 20.74  (0.19) & 18.90  (0.03) & 19.96  (0.04) & 17.97  (0.02) & 17.59  (0.03) & 17.37  (0.04) & NOT \\

\hline\hline 
\tabularnewline[-0.25cm]
\end{tabular}
\caption{Journal detailing the photometric follow-up campaign. Images were taken at the Liverpool Telescope (LT) using IO:O and the Nordic Optical Telescope (NOT) using ALFOSC, situated at the Roque de los Muchachos Observatory in La Palma. The epoch is given relative to the presumed explosion date, MJD=58208.0.} 
\label{phot}  
\end{sidewaystable}

\begin{figure}
   \centering
   \includegraphics[width=\hsize]{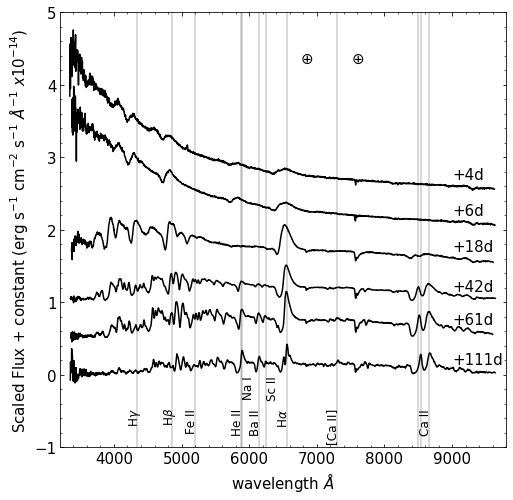}
      \caption{Spectral evolution of SN~2018aoq over 4 months. The spectra have been offset for clarity and corrected for both redshift and extinction.}
         \label{spectra}
   \end{figure}

\begin{table}
\begin{tabular}{l c r c c}        
\hline\hline                 
\tabularnewline[-0.25cm]
Date & MJD & Epoch & Filter & Magnitude\\
 & (+2400000) & (d)& & \tabularnewline
\hline                        
\tabularnewline[-0.25cm]

2018 Mar 11 & 58188.50 & -19.50 & o & >18.06 \\
2018 Mar 16 & 58194.46 & -13.54 & c & >20.37 \\
2018 Mar 19 & 58196.50 & -11.50 & c & >19.87 \\
2018 Mar 28 & 58206.47 & -1.53 & o & >18.89 \\
2018 Apr 05 & 58214.47 & 6.47 & o & 15.30 (0.01)\\
2018 Apr 11 & 58220.45 & 12.45 & o & 15.34 (0.01)\\
2018 Apr 14 & 58222.53 & 14.53 & o & 15.33 (0.01)\\
2018 Apr 15 & 58224.46 & 16.46 & c & 15.56 (0.02)\\
2018 Apr 23 & 58232.43 & 24.43 & o & 15.41 (0.01)\\
2018 Apr 25 & 58234.39 & 26.39 & o & >15.36 \\
2018 Apr 25 & 58234.41 & 26.41 & o & 15.33 (0.32)\\
2018 May 05 & 58244.41 & 36.41 & o & 15.46 (0.02)\\
2018 May 07 & 58246.42 & 38.42 & o & 15.35 (0.01)\\
2018 May 09 & 58248.41 & 40.41 & o & 15.40 (0.01)\\
2018 May 11 & 58250.40 & 42.40 & o & 15.33 (0.01)\\
2018 May 13 & 58251.51 & 43.51 & c & 15.68 (0.01)\\
2018 May 13 & 58252.48 & 44.48 & c & 15.67 (0.01)\\
2018 May 15 & 58254.42 & 46.42 & o & 15.40 (0.01)\\
2018 May 17 & 58256.40 & 48.40 & c & 15.66 (0.01)\\
2018 May 19 & 58258.43 & 50.43 & o & 15.35 (0.01)\\
2018 May 21 & 58260.43 & 52.43 & o & 15.42 (0.06)\\
2018 May 23 & 58262.36 & 54.36 & o & 15.53 (0.27)\\
2018 May 25 & 58264.37 & 56.37 & o & 15.36 (0.02)\\
2018 May 27 & 58266.36 & 58.36 & o & 15.34 (0.05)\\
2018 May 31 & 58270.34 & 62.34 & o & 15.37 (0.04)\\
2018 Jun 02 & 58272.34 & 64.34 & o & 15.35 (0.01)\\
2018 Jun 04 & 58274.33 & 66.33 & o & 15.32 (0.01)\\
2018 Jun 06 & 58276.32 & 68.32 & o & 15.32 (0.02)\\
2018 Jun 08 & 58278.31 & 70.31 & o & 15.32 (0.01)\\
2018 Jun 10 & 58280.34 & 72.34 & c & 16.36 (0.02)\\
2018 Jun 12 & 58282.30 & 74.30 & o & 15.36 (0.01)\\
2018 Jun 14 & 58284.33 & 76.33 & c & 16.39 (0.02)\\
2018 Jun 16 & 58286.28 & 78.28 & o & 15.36 (0.01)\\
2018 Jun 18 & 58288.28 & 80.28 & o & 15.41 (0.01)\\
2018 Jun 20 & 58290.27 & 82.27 & o & 15.53 (0.01)\\
2018 Jun 22 & 58292.27 & 84.27 & o & 15.52 (0.02)\\
2018 Jun 24 & 58294.27 & 86.27 & o & 15.56 (0.02)\\
2018 Jun 26 & 58296.29 & 88.29 & o & 15.55 (0.01)\\
2018 Jun 27 & 58297.32 & 89.32 & o & 15.62 (0.04)\\
2018 Jun 28 & 58298.26 & 90.26 & o & 15.63 (0.02)\\
2018 Jun 30 & 58300.27 & 92.27 & o & 15.69 (0.01)\\
2018 Jul 02 & 58302.28 & 94.28 & o & 15.67 (0.01)\\
2018 Jul 04 & 58304.26 & 96.26 & o & 15.70 (0.05)\\
2018 Jul 10 & 58310.26 & 102.26 & o & 15.79 (0.06)\\
2018 Jul 12 & 58312.28 & 104.28 & c & 18.05 (0.13)\\
2018 Jul 16 & 58316.28 & 108.28 & c & >20.13 \\
2018 Jul 18 & 58318.26 & 110.26 & o & 16.14 (0.04)\\
2018 Jul 20 & 58320.26 & 112.26 & o & 16.96 (0.19)\\
2018 Jul 22 & 58322.26 & 114.26 & o & 16.67 (0.05)\\
2018 Jul 24 & 58324.26 & 116.26 & o & 17.29 (0.07)\\
2018 Jul 30 & 58330.25 & 122.25 & o & 17.66 (0.13)\\
2018 Jul 31 & 58331.30 & 123.30 & c & >16.56 \\
2018 Aug 07 & 58338.24 & 130.24 & o & 17.98 (0.38)\\
2018 Aug 13 & 58344.25 & 136.25 & c & >18.00 \\
\hline\hline 
\tabularnewline[-0.25cm]
\end{tabular}
\caption{Photometric journal detailing the ATLAS observations of SN~2018aoq. Measurements from images taken in the same filter and on the same night were averaged. } 
\label{ATLAStable}  
\end{table}
\begin{table*}
\begin{center}
\begin{tabular}{c c c c c c c}        
\hline\hline                 
\tabularnewline[-0.25cm]
Date & MJD& Phase & Instrument  & Filter & Exposure time & Vega mag.\\
 & (+2400000)  & (d) & & & (s) \tabularnewline
\hline                        
\tabularnewline[-0.25cm]

2015 Dec 07 & 57363.06 &-845& WFC3/UVIS & F350LP & 1050&25.67 (0.06)\\
2015 Dec 07 & 57363.08 &-845& WFC3/IR & F160W & 1106&21.79 (0.02)\\
2015 Dec 14 & 57370.68 &-837& WFC3/UVIS & F350LP & 1050&25.69 (0.04)\\
2015 Dec 14 & 57370.70 &-837& WFC3/UVIS & F555W & 1100&26.61 (0.11)\\
2015 Dec 21 & 57377.92 &-830& WFC3/UVIS & F814W & 1100&23.77 (0.04)\\
2015 Dec 28 & 57384.74 &-823& WFC3/IR & F160W & 1106&21.89 (0.02)\\
2016 Jan 03 & 57390.02 &-818& WFC3/UVIS & F350LP & 1050&25.83 (0.05)\\
2016 Jan 03 & 57390.04 &-818& WFC3/UVIS & F555W & 1100&26.77 (0.12)\\
2016 Jan 05 & 57392.77 &-815& WFC3/UVIS & F350LP & 1050&25.69 (0.04)\\
2016 Jan 12 & 57399.14 &-809& WFC3/UVIS & F814W & 1100&24.10 (0.05)\\
2016 Feb 05 & 57423.56 &-784& WFC3/UVIS & F350LP & 1050&25.81 (0.04)\\
2016 Feb 11 & 57429.54 &-778& WFC3/UVIS & F814W & 1100&24.07 (0.04)\\
2016 Feb 18 & 57436.89 &-771& WFC3/UVIS & F350LP & 1050&25.76 (0.04)\\
2018 Apr 23* & 58232.17 &+24& WFC3/UVIS & F555W & 270&15.79 (0.01)\\
\hline\hline 
\tabularnewline[-0.25cm]
\end{tabular}
\caption{Record of the archival WFC3 images used to identify the progenitor.
*This image contains the SN (see proposal 15151).} 
\label{Archive}  
\end{center}
\end{table*}
\begin{table}
\begin{tabular}{l c r c}        
\hline\hline                 
\tabularnewline[-0.25cm]
Date & MJD & Phase & Exposure Time\\
 & (+2400000) & (d)& (s)\tabularnewline
\hline                        
\tabularnewline[-0.25cm]

2018 Apr 03 & 58211.64 & 4 & 1200 \\
2018 Apr 05 & 58213.61 & 6 & 1200 \\
2018 Apr 17 & 58225.51 & 18 & 1200 \\
2018 May 11 & 58249.52 & 42 & 1200 \\
2018 May 30 & 58269.43 & 61 & 1200 \\
2018 Jul 19 & 58319.39 & 111 & 1800 \\

\hline\hline 
\tabularnewline[-0.25cm]
\end{tabular}
\caption{Details of spectroscopic observations of SN~2018aoq. All spectra were taken using the Nordic Optical Telescope ALFOSC as part of the NUTS programme. Grism 4 (3200-9600\AA) was used in a combination with a $1\farcs0$ slit providing a resolution of 16.2\,\AA. The second order contamination of grism4 was corrected using the prescriptions of \cite{g4cor}. } 
\label{spectable}  
\end{table}

\begin{table*}
        \begin{tabular}{c c c c c c c c c c}        
        \hline\hline                 
        \tabularnewline[-0.25cm]
No. & RA & DEC & $u$ & $B$ & $V$ & $g$ & $r$ & $i$ & $z$ \tabularnewline
\hline                        
\tabularnewline[-0.25cm]
1 & 12:10:38.604 & +39:20:29.725 & 16.17 & 15.60 & 14.46 & 15.02 & 14.08 & 12.19 & 12.61\\
2 & 12:10:16.845 & +39:19:24.478 & 17.98 & 16.67 & 15.80 & 16.20 & 15.54 & 15.32 & 15.22\\
3 & 12:10:12.437 & +39:19:34.249 & 14.83 & 13.73 & 12.83 & 13.25 & 12.56 & 12.43 & 13.08\\
4 & 12:10:23.596 & +39:21:24.185 & 16.90 & 15.81 & 15.08 & 15.40 & 14.88 & 14.72 & 14.63\\
5 & 12:10:25.285 & +39:23:55.432 & 16.17 & 15.11 & 14.39 & 14.70 & 14.19 & 14.07 & 14.02\\
6 & 12:10:30.733 & +39:25:36.616 & 15.51 & 14.17 & 14.58 & 14.21 & 14.85 & 16.14 & 13.55\\
7 & 12:10:37.775 & +39:25:09.376 & 20.50 & 18.51 & 17.21 & 17.87 & 16.76 & 16.35 & 16.12\\
8 & 12:10:39.930 & +39:23:52.103 & 18.76 & 18.18 & 17.68 & 17.86 & 17.58 & 17.47 & 17.44\\
9 & 12:10:48.489 & +39:26:54.815 & 22.84 & 20.82 & 19.14 & 20.03 & 18.54 & 17.21 & 16.51\\
10& 12:10:55.625 & +39:26:24.242 & 18.34 & 17.50 & 16.87 & 17.13 & 16.71 & 16.54 & 16.47\\
\hline                                   
        \end{tabular}
    \caption{Photometry of the reference stars as labelled in figure \ref{image}. }  
    \label{Cal}  
\end{table*}

\begin{table}
        \begin{tabular}{c c c}        
        \hline\hline                 
        \tabularnewline[-0.25cm]
SN & $E(B-V)$ & Reference \tabularnewline
\hline                        
\tabularnewline[-0.25cm]
2005cs & 0.050 & \cite{Baron05cs}\\
1999em & 0.100 & \cite{Hamuy99em}\\
1999br & 0.024 & NED, \cite{PasLL}\\
2009md & 0.100 & \cite{09md}\\
2003gd & 0.100 & \cite{03gdEBV}\\
2004eg & 0.399 & \cite{Spiro14}\\
2008in & 0.098 & \cite{2008in}\\
2004dj & 0.100 & \cite{Vinko04dj}\\
\hline                                   
        \end{tabular}
    \caption{II-P sample used to calculate extinction.}  
    \label{ExSample}  
\end{table}

\end{appendix}

\end{document}